\documentclass[onecollarge,natbib]{svjour2}
\bibpunct{[}{]}{;}{n}{}{,} 
\smartqed  
\usepackage{graphicx}
\newcommand{\Hpart}{\textsc H}
\journalname{Few-Body Syst}
\usepackage{amsmath}
\usepackage{amssymb}
\usepackage{xcolor}
\usepackage{tensor}
\usepackage{braket}
\usepackage{units}
\usepackage[hypertex,colorlinks=true,linkcolor=red,citecolor=blue]{hyperref}

\begin{document}

\title{Covariance constraints for light front wave functions}

\author{D.~M\"uller}


\institute{D.~M\"uller \at
              Theoretical Physics Division, Rudjer Bo{\v s}kovi{\'c} Institute, HR-10000 Zagreb, Croatia\\
              \email{dieter.mueller@irb.hr}           
}

\date{Received: date / Accepted: date}

\maketitle

\begin{abstract}
Light front wave functions (LFWFs) are often utilized to model parton distributions and form factors where their transverse and longitudinal momenta are
tied to each other in some manner that  is often guided by convenience. On the other hand, the cross talk of  transverse and longitudinal momenta is
governed by Poincar{\'e} symmetry and thus popular LFWF models are often not usable to model more intricate quantities such
as  generalized parton distributions. In this contribution  a closer look to this issue is given and it is shown how to overcome the issue for  two--body LFWFs.

\end{abstract}

\section{Introduction}
\label{intro}

In Quantum Chromodynamics (QCD) inspired phenomenology it is popular to discuss the partonic picture of observables, such as form factors, structure functions of inclusive processes, spectrography, and many other quantities as well as  partonic quantities, defined as matrix elements of   two-body operators, such as distribution amplitude, parton distribution functions, generalized parton distributions (GPDs), transverse momentum dependent parton distributions, and  phase space functions in terms of LFWFs \cite{DreYan69,Brodsky:1997de,Boffi:2007yc}.  Unfortunately, LFWFs could so far not be determined from the QCD dynamics. Since in a Hamilton approach the underlying Poincar{\'e} symmetry is not explicitly manifested, it arises for model builders the problem how in the LFWFs the longitudinal and transverse momenta are tied to each other. If one considers translation invariant parton distributions or hadronic distribution amplitudes, this issue does not imply {\em visible} inconsistencies, however, it becomes crucial if one deals with non-forward quantities such as generalized parton distribution or phase space functions.

In particular, for the phenomenology of  deeply virtual Compton scattering \cite{Mueller:1998fv,Radyushkin:1996nd,Ji:1996nm} and
deeply virtual meson production \cite{Radyushkin:1996ru,Collins:1996fb} the Poincar{\'e} covariance of GPDs is crucial.  In the double distribution (DD) representation \cite{Mueller:1998fv,Radyushkin:1997ki} and the double partial wave expansion, see, e.g.,  \cite{Muller:2014wxa} and references therein, this property is manifestly implemented, however, if one likes to model GPDs in terms of LFWFs this might be considered as a theoretical challenge. In the LFWF overlap representation \cite{Brodsky:2000xy,Diehl:2000xz}  the outer GPD region in the momentum fraction $x$ and skewness $\eta$ plane, i.e. $|\eta| \le |x|\le 1$, is obtained from the parton number conserved overlap while the central GPD region i.e. $ |x| \le |\eta|$, arises from the parton number changing LFWF overlap. The GPD covariance condition  ensures that the GPD moments 
\begin{equation}
\label{GPD2moments}
\int_{-1}^{1}\!dx\, x^n H(x,\eta,t) = \sum_{m=0 \atop m\ {\rm even}}^{n+1}H_{nm}(t)  \eta^m \,,
\end{equation}
where $t$  is the momentum transfer square, are even polynomials in $\eta$ of order  $n$ (even $n$) or $n+1$ (odd  $n$).
Moreover, covariance also ensures that the amplitudes, given as convolution of partonic amplitudes and  GPDs, satisfy  fixed-$t$ dispersion relations \cite{Teryaev:2005uj}, e.g., at leading order accuracy they are written as 
 \begin{equation}
\label{H-DR}
{\cal H }^\sigma(\xi, \vartheta,t) \stackrel{\rm LO}{=} \int_{0}^1\!dx \frac{(1+\sigma)x + (1-\sigma) \xi}{\xi^2-x^2}H^{\sigma}(x,\eta=\vartheta x,t) + \frac{1+\sigma}{2}{\cal D}(\vartheta,t)\,,
\end{equation}
where $\xi$ is a Bjorken-like scaling variable, $\vartheta=\eta/\xi$ is the asymmetry parameter, $\sigma =\pm 1$ is the signature factor, and  ${\cal D}(\vartheta,t)$ is a subtraction constant, called $D$-term form factor.
However, for LFWFs that are ambiguously  modeled as functions of longitudinal and transverse momenta it is very likely that the covariance property of GPD and so the analyticity of the amplitudes is violated. Also in many applications only the outer GPD region could be calculated from the parton conserved  LFWF overlap and it is sometimes believed that the (unknown) non-conserved parton number LFWF overlap contribution will render GPDs that are covariant.

On the other hand, the Poincar{\'e} covariance property ensures that the outer region, i.e., the partonic $s$--channel view, and the central  one, i.e., the partonic $t$--channel view, are dual to each other. More precisely, the central region can be uniquely mapped to the outer one \cite{Mueller:2005ed} (this map was first applied for evolution kernels in \cite{Dittes:1988xz}), while the inverse map is only unique in the charge odd sector, however, in the charge even sector a so-called $D$-term contribution \cite{Polyakov:1999gs}, which entirely lives in the central region and vanishes on the cross--over line $|x|=|\eta|$, might appear. One might fix this contribution by some external requirements, however, no proof on general ground is known \cite{Muller:2015vha}. Thus, one might be able to consistently model the  GPDs in terms of parton number conserved overlap of LFWFs, where duality is used to restore the full GPD. The condition, which is left, is on the functional form of LFWFs, i.e., how longitudinal and transverse momenta are tied to each other. The advantage of such a framework is besides a direct interpretation of  experimental measurements in terms LFWFs also that positivity constraints, which are in their most general formulation are given impact parameter space \cite{Pobylitsa:2002iu}, should be manifestly  implemented in the LFWF overlap representation.


\section{GPD duality}
\label{sec:duality}

\label{eq:defGPDV}
The GPD support properties, given here within the restriction $|x|\le 1$, can be derived from the DD representation and might be written in terms of quark ($q$) and anti-quark ($\overline{q}$)  building blocks
\begin{eqnarray}
H(x, \eta, t) &=& -\theta(x\le-|\eta|) \left[ \Hpart^{\overline{q}}(-x,\eta,t) + \Hpart^{\overline{q}}(-x,-\eta,t)\right]
+
\theta(x\ge|\eta|)\left[ \Hpart^q(x,\eta,t) + \Hpart^q(x,-\eta,t)\right]
\nonumber\\
&&+
\theta(|x|\le|\eta|)\left[-\Hpart^{\overline{q}}(-x,\eta,t)  + \Hpart^q(x,\eta, t)\right],
\label{eq:GPDsupport}
\end{eqnarray}
where the building block for the (anti-)quark GPD has the integral representation
\begin{equation}
\label{eq:GPDbb}
\Hpart^i(x,\eta,t) = \frac{1}{\eta}\int_{0}^{\frac{x+\eta}{1+\eta}}\!dy\, h(y,(x-y)/\eta,t)\,, \quad  i\in\{u, \bar{u}, d, \bar{d},\cdots\}\,.
\end{equation}
Note that the DD $h(y,z,t)$, which has the symmetry property $h(y,-z,t)= h(y,z,t)$, might be undefined in the outer region for $y< \frac{x-\eta}{1-\eta}$, however, the GPD in this region is defined,
\begin{equation}
\label{eq:GPDout}
H(x\ge \eta, \eta, t)=
\Hpart^i(x,\eta,t) + \Hpart^i(x,-\eta,t) =  \frac{1}{\eta}\int_{\frac{x-\eta}{1-\eta}}^{\frac{x+\eta}{1+\eta}}\!dy\, h(y,(x-y)/\eta,t)  \,.
\end{equation}
Formulae (\ref{eq:GPDsupport})  and (\ref{eq:GPDbb}) ensure that the $x$-moments of the GPD are  polynomials (\ref{GPD2moments}) and that amplitudes are calculable from (\ref{H-DR}).
Utilizing the dispersive framework and the operator product expansion for doubly deep virtual Compton scattering in the Euclidean region \cite{Kumericki:2007sa,Kumericki:2008di} yields for  $|\vartheta| \le 1$ the equality
\begin{equation}
\label{DDVCS-equality}
{\rm PV}\!\int_{-1}^1\!dx \frac{1}{\xi-x}
H^{\sigma}(x,\eta,t)
={\rm PV}\!\int_{0}^1\!dx \frac{(1+\sigma)x + (1-\sigma) \xi}{\xi^2-x^2}H^{\sigma}(x,\eta=\vartheta x,t)+ \frac{1+\sigma}{2}  {\cal D}(\vartheta,t) \,.
\end{equation}

As annulled in the Introduction, we are interested to construct the full GPD from the outer region and so far various analytic procedures that are related to each other were suggested:
\begin{enumerate}
\item The truncated Taylor expansion of the moments  $\int_{\eta}^1\!dx\, x^n H(x,\eta,t)$, where a possible non-analytic part must be removed first,
 yields even polynomials in $\eta$  of order $n$ or $n+1$, see Ref.\ \cite{Mueller:2005ed},
 \begin{equation}
 \label{H^{trun}_n}
 H^{\rm trun}_n(\eta,t|\sigma) = \sum_{m=0}^{n+1}\frac{1}{m!} \frac{d^m}{d\eta^m}\left\{\int_\eta^1\!dx\,x^n\left[ H(x,\eta,t) -\sigma H(-x,\eta,t)\right]-
 \mbox{non-analytic part}\right\}\Big|_{\eta=0}.
\end{equation}
The analytic continuation of odd $n$ ($\sigma=+1$) and even $n$ ($\sigma=-1$)  polynomials and an inverse Mellin transform allows to find the  GPDs with definite signature  $H^{\sigma} =H(x,\eta,t) - \sigma H(-x,\eta,t)  $,
\begin{equation}
\label{Hn2H}
H^\sigma(x\ge 0,\eta,t) =\frac{1}{2\pi i} \int_{c-i \infty}^{c+i \infty}\!dx\, x^{-n-1} H^{\rm trun}_{n}(\eta,t|\sigma) + \frac{1+\sigma}{2} \theta(0\le x\le |\eta|)2\, \delta\! D(x/|\eta|,t)\,.
\end{equation}
These GPDs $H^{\pm}$ have the support $0\le x\le 1$, their continuation to negative $x$ is done by antisymmetrization or symmetrization in $x$. For a signature even (or charge even) GPD it is allowed to add an additional $\delta\! D(z,t)=\delta\! D(-z,t)$ with $\delta\! D(\pm 1,t)=0$   contribution to the intrinsic $D$-term.
\item The GPD on the r.h.s.\ of (\ref{DDVCS-equality}), i.e., $H^\sigma(x,\eta=\vartheta x)$, is only scanned in the outer GPD region and the l.h.s.\  can be inverted by a Hilbert transform or under certain assumptions by an alternative integral transform over the region 
$\xi \in [-1,1]$ \cite{Kumericki:2008di}, providing the full GPD for definite signature
\begin{equation}
\label{H-Hilbert}
H^\sigma(x,\eta,t) = \frac{{\rm PV}}{\pi^2} \!\! \int_{-\infty}^\infty\!d\xi \frac{\Re{\rm e}{
\cal H }^\sigma(\xi,\eta/\xi,t)}{\xi-x} 
\;\mbox{or}\;
H^\sigma(x,\eta,t) = \frac{{\rm PV}}{\pi^2} \!\! \int_{-1}^1\!d\xi \frac{\sqrt{1-x^2}}{\sqrt{1-\xi ^2}}\frac{\Re{\rm e}{
\cal H }^\sigma(\xi,\eta/\xi,t)}{\xi-x},
\end{equation}
where ${\cal H }^\sigma$, calculated from the dispersion integral (\ref{H-DR}), must be continued into the region $|\vartheta| \ge 1$.

\item Expressing the GPD in the outer region by the DD--integral (\ref{eq:GPDout}) allows to read off the DD itself, i.e., the  complete GPD  can be restored \cite{Hwang:2007tb,Muller:2014tqa}, for an example see next section.
\end{enumerate}

To exemplify the first two methods let us consider the photon GPD, obtained from a one-loop calculation \cite{Friot:2006mm}, which reads up to a normalization factor in the outer region as following
\begin{eqnarray}
\label{H_1-outer}
H_1(x\ge \eta,\eta) =  1-\frac{2x(1-x)}{1-\eta ^2}\,.
\end{eqnarray}
The truncated moments (\ref{H^{trun}_n}) can be straightforwardly calculated and are even polynomials in $\eta$ of order $n$ (even $n$) or $n+1$ (odd $n$).  With $\sigma=-(-1)^n$ they can be written in terms of a rational function
\begin{eqnarray}
H^{\rm trun}_{1,n}(\eta|\sigma) =
\frac{2-\left[1+\sigma\right] \eta^{1+n}}{2(n+1)}
-\frac{2-\left[1+\sigma\right] \eta ^{3+n}-\left[1-\sigma\right]\eta ^{2+n}}{(2+n) (3+n) \left(1-\eta ^2\right)}\, .
\end{eqnarray}
Apart from the highest possible power $\eta^{n+1}$ for odd $n$ these polynomials coincide with those of the originally calculated GPD, see Eq.\ (23) in \cite{Mueller:2014hsa}.  If the $D$-term related $\eta^{n+1}$ coefficient is known, the GPD can uniquely restored in the central region. In our example the addenda $\delta\!D_n = \frac{\eta^{n + 1}}{n + 2}$ should be added and the inverse Mellin transform (\ref{Hn2H}) yields for $\sigma=+1$ the original GPD, written here as
 \begin{eqnarray}
\label{H1-full}
H_{1}(x\ge 0,\eta) &=& \frac{1}{2\pi i} \int_{c-i \infty}^{c+i \infty}\!dx\, x^{-n-1} \left[ H^{\rm trun}_{1,n}(\eta|\sigma=+1) + \delta\!D_n\right],
 \quad \delta\!D_n = \frac{\eta^{n + 1}}{n + 2}
\\
&=& \theta(0\le x\le \eta)\, \frac{x (1-\eta )}{\eta  (1+\eta )} + \theta(\eta\le x\le 1)\left[1-\frac{2 (1-x) x}{1-\eta ^2}\right] \quad\mbox{for}\quad
\eta \ge 0\,.
\nonumber
\end{eqnarray}

To employ the second method, we first calculate the $\sigma=+1$ amplitude from the  dispersion relation (\ref{H-DR}), where the 
$D$-term form factor, considered as known, is
evaluated from the $D$-term (26) of \cite{Mueller:2014hsa},
\begin{eqnarray}
{\cal H}_1(\xi,\vartheta) &=&
\frac{2 \vartheta  \ln\!\frac{1+\vartheta}{1-\vartheta }+2\ln\!\left(1-\vartheta ^2\right)}{\vartheta^2 \left(1-\vartheta ^2 \xi ^2\right)}+\frac{ \left[1+(2-\vartheta^2) \xi ^2\right]\ln\!\frac{\xi ^2}{\xi^2-1}-2 \xi \ln\!\frac{1+\xi}{\xi-1}}{ 1-\vartheta ^2 \xi ^2} +  {\cal D}(\vartheta),
\\
{\cal D}(\vartheta)  &=& \int^{\vartheta}_{- \vartheta}\!dx  \frac{2x\,D(x, \vartheta)}{1-x^2}\,= -\frac{\vartheta  \ln \frac{1+\vartheta }{1-\vartheta }+\left(2-\vartheta ^2\right) \ln(1-\vartheta ^2)}{\vartheta ^2}.
\nonumber
\end{eqnarray}
Performing the Hilbert transform (\ref{H-Hilbert}) yields the GPD (\ref{H1-full}). Note that the GPD  can be alternatively calculated from the imaginary part of the amplitude, i.e., $H_1(x,\eta)=\Im{\rm m} {\cal H}_1(x-i \epsilon, \eta/[x-i \epsilon])/\pi$.

Let us emphasize that an approximate restoration of polynomiality yields incorrect results for the real part of amplitudes if they are calculated from the GPD convolution formula. That this effect can be large in the presence of Regge-like behavior has been exemplified in \cite{Diehl:2007jb}.

\section{Functional form of two-body LFWFs}
\label{sec:LFWFs}

The restoration of the full GPD from the parton number conserved LFWF overlap, i.e., from the outer GPD region, is only possible if the LFWFs respect the underlying Poincar{\'e} symmetry.  There are various suggestions in the literature how covariance might be implemented in LFWFs \cite{Brodsky:1980vj,Brodsky:2003pw,Muller:2014tqa}. The  GPD covariance property has two aspects:
\begin{itemize}
\item restoration of $t$-dependence from the two-dimensional transverse momentum vector ${\bf \Delta}_\perp=(\Delta_x,\Delta_y)$
\item ensuring that the resulting GPDs satisfy  the polynomiality condition.
\end{itemize}
In Ref.\ \cite{Muller:2014tqa} both of these problems were solved for effective two-body LFWFs. The restoration of $t$ dependence can be ensured by a
$({\bf k}_\perp^2 - X(1-X) M^2)/(1-X)$  dependence and the GPD polynomiality constraints  provide a further restriction on the functional form of LFWFs, which might be
conveniently written  as a Laplace transform
\begin{equation}
\label{LFW-ansatz}
 \phi(X,{\bf k}_\perp) = \int_0^\infty\! d\alpha\; \varphi(\alpha) \exp\left\{-\alpha  \frac{{\bf k}_\perp^2 + (1-X) m^2 + X \lambda^2 - X(1-X) M^2}{(1-X) M^2}  \right\},
\end{equation}
where $X$ and $m$ is the longitudinal momentum fraction and the mass of the struck quark, $\lambda$ is the spectator quark mass, $M$ is the hadron mass,  ${\bf k}_\perp$ is the transverse momentum, and $\varphi(\alpha)$ might be somehow interpreted as a reduced wave function. The transverse and longitudinal momenta are tied to each other by the off-shell propagator \cite{Brodsky:1980vj}
$$M^2 - k_1^- -k_2^- =M^2- \frac{{ \bf k}_\perp^2 + m^2}{X} - \frac{{ \bf k}_\perp^2 + \lambda^2}{1-X} ,$$
which, however, in (\ref{LFW-ansatz}) is additionally scaled by the momentum fraction $X$.
The ${\bf k}_{\perp}$-unintegrated LFWF overlap or unintegrated GPD in the outer region is defined in such models as
\begin{equation}
\label{GPD-overlap}
{\bf H}(x \ge \eta,\eta,{\bf \Delta}_{\perp},{\bf k}_{\perp}) =
\frac{1}{1-x}
\phi^\ast\!\left(\frac{x-\eta}{1-\eta},{\bf k}_{\perp}- \frac{1-x}{1-\eta} {\bf \Delta}_\perp\right)
\phi\!\left(\frac{x+\eta}{1+\eta},{\bf k}_{\perp}\right),
\end{equation}
where for simplicity we do not discuss here the quark and hadron spin content. For $x\ge \eta$ this GPD
can be equivalently written in terms of an unintegrated DD representation \cite{Muller:2014tqa},
\begin{equation}
\label{Def-DDunint}
{\bf H}(x,\eta,{\bf
\Delta}_{\perp},{\bf
k}_{\perp}) = \int_0^1\! dy\int_{-1+y}^{1-y}\! dz\;  \delta(x-y-z \eta)\,
 {\bf h}\!\left(y,z,t,\overline{\bf k}_\perp\!\right),
\end{equation}
which allows us to extend the support into the central region $-\eta \le x \le \eta $.
This unintegrated DD can be expressed after some algebra in terms of the LFWF (\ref{LFW-ansatz}),
\begin{eqnarray}
\label{Def-DD-unint1}
{\bf  h}(y,z,t,\overline{{\bf k}}_\perp) &=& \frac{1}{2}\int_0^\infty\!dA\,A\;
\varphi^\ast\!\left(\!A \frac{1-y+z}{2}\!\right) \varphi\!\left(\!A \frac{1-y-z}{2}\!\right)
\\
&&\times
\exp \left\{-A\left[
(1-y)\frac{m^2}{M^2}+y \frac{\lambda^2}{M^2}-y(1-y)-\left[(1-y)^2-z^2\right] \frac{t}{4 M^2}+\frac{\overline{{\bf k}}^2_\perp}{M^2}
\right] \right\},\nonumber
\end{eqnarray}
depending on $\overline{{\bf k}}_\perp={\bf k}_\perp-(1-y+z) {\bf  \Delta}_\perp/2$, the set $\{m,\lambda, M\}$ of mass parameters, and the reduced LFWF $\varphi(\alpha)$. The integration over
the transverse degrees of freedom can trivially be performed and we recover from (\ref{Def-DDunint}) the common DD--representation for a `quark' GPD,
\begin{equation}
\label{Def-DD-unint2}
H(x,\eta,t) =  \int_0^1\! dy\int_{-1+y}^{1-y}\! dz\;  \delta(x-y-z \eta)\, h(y,z,t)\,,  \quad
h(y,z,t) = \int\!\!\!\! \int\!d^2\overline{{\bf k}}_\perp\, {\bf  h}(y,z,t,\overline{{\bf k}}_\perp) \,.
\end{equation}
which manifestly implements the GPD covariance properties. 
Two comments are in order.
\begin{itemize}
\item GPDs build with two-body LFWF models possess a cross-talk among $t$- and $\eta$-dependence.
\item Other functional forms of two-body LFWFs  than (\ref{LFW-ansatz})  might be not capable to deliver a GPD that respects covariance and, consequently, analyticity, i.e., the equality (\ref{DDVCS-equality}), can not be satisfied.
\end{itemize}
Let us exemplify these points, where for simplicity  the mass parameters $M/2=\lambda = m$ are equated and  $\varphi(\alpha)=\delta(\alpha - \alpha_0)$ is taken in the ansatz (\ref{LFW-ansatz}).  Plugging the resulting exponential LFWF
\begin{equation}
\phi(x,{\bf k}_\perp) =  \exp\left\{-\alpha_0 \frac{{\bf k}_\perp^2+  m^2 - 4X(1-X) m^2 }{4(1-X) m^2}  \right\}
\end{equation}
 in the overlap representation  (\ref{GPD-overlap}) and performing the ${\bf k}_\perp$--integration yields a  GPD in the other region,
\begin{equation}
{H}(x\ge \eta,\eta,t) \propto  \exp\left\{-\frac{\alpha_0}{2}\left[ \frac{ (1-2 x)^2}{1-x} - \frac{(1-x)t }{4m^2}\right]\right\},
\end{equation}
that is independent on $\eta$.
Thus the amplitude  (\ref{H-DR})  is $\vartheta$--independent and the Hilbert transform  (\ref{H-Hilbert}) tells us that the extension into the central region is done by replacing
the constraint $x\ge \eta$ by $x\ge 0$.
Of course, this result can be also obtained from the DD representation (\ref{Def-DD-unint1}) and (\ref{Def-DD-unint2}) .

Contrarily, the construction of a consistent GPD might be impossible if we take an arbitrary functional form. For instance, picking up one of the popular choices for modeling meson LFWFs \cite{Huang:2013yya},
\begin{equation}
\phi(x,{\bf k}_\perp) =  \exp\left\{-\alpha_0 \frac{{\bf k}_\perp^2+  m^2 - 4X(1-X) m^2 }{4(1-X) X m^2}  \right\},
\end{equation}
we immediately get from the overlap  formula (\ref{GPD-overlap})  the GPD in the outer region
\begin{equation}
{H}(x\ge \eta,\eta,t) \propto\frac{x^2-\eta ^2}{x \left(1+\eta ^2\right)-2 \eta ^2}
 \exp\!\!{\left\{\!\!-\frac{\alpha_0 (1-\eta ^2)/2}{x (1+\eta ^2)-2 \eta ^2}\!\!
  \left[\!\!
  \frac{x^2 \left(1+\eta ^2-2 x\right)^2}{ (1-x)\left(1-\eta ^2\right) \left(x^2-\eta ^2\right)}- \frac{(1-x)t}{4 m^2}
  \!\right]\!\!\right\}}.
\end{equation}
Utilizing the first method of the preceding section, one realizes that
the coefficients in the truncated Taylor expansion (\ref{H^{trun}_n}) suffer  from essential singularities, except for the constant term. Thus, removing these singularities yields to a $\eta$-independent GPD. Also the  second method fails, namely, the ``spectral function'' takes, e.g., for $t=0$, the form
\begin{equation}
{H}(x,\eta=\vartheta x,t=0) \propto \frac{ x \left(1-\vartheta ^2\right)}{1-x(2-x) \vartheta ^2}
 \exp\!\left\{-\frac{\alpha_0}{2} \frac{ \left[1-x(2-x \vartheta ^2)\right]^2}{ (1-x) x \left(1-\vartheta ^2\right) \left[1-x(2-x) \vartheta ^2\right]}
 \right\}.
\end{equation}
and  has for $0\le x\le 1$  essential singularities at $\vartheta^2 =1$ and $\vartheta^2=1/x(2-x)$.  Consequently,
the resulting amplitude (\ref{H-DR}) does not exist for $\vartheta^2 > 1$ and its limiting value at $\vartheta^2=1$ depends on the direction.    It is not hard
to realize that more general exponential LFWF ans\"atze are also not usable to build  GPDs.
\vspace{-0.15cm}

\section{Conclusions}

We recalled three methods that allow to restore the full GPD from the knowledge of the GPD in the outer region. All these methods are based on extension procedures. So far they can be only utilized if the GPD in the outer region is known analytically. The extension of GPDs is tied to the analytic properties of amplitudes which besides the dispersion relation representation with respect to the Bjorken-like scaling variable $\xi$ must also possess certain  properties with respect to the photon asymmetry parameter $\vartheta$, namely, analyticity inside the unit circle and extendable on the segment $\vartheta \in [1,\infty]$ (the value on $[-\infty,-1]$ follows from symmetry requirements). The investigation of the remaining mathematical problems is important if one likes to have a numerical GPD extension procedure, which can then be conveniently employed in GPD phenomenology. Without relying on an external principle,  the freedom which is left in the charge even sector is the so-called $D$-term,  entirely living in the central GPD region.

From the partonic point of view it is tempting to interpret experimental measurements of deeply virtual Compton scattering and deeply virtual meson production in terms of (effective) LFWFs.
Such a framework also offers the possibility to implement positivity constraints in the phenomenological analysis, however, here it is requested that the resulting amplitudes possess the correct analytic properties.  It is expected that this is ensured if the  LFWFs satisfy the Poincar{\'e} covariance constraints and so one can utilize GPD duality to restore (apart from the $D$-term) the full GPD in the central region, too.  It was exemplified here within  exponential LFWFs, used for mesons, that  if such constraints are neglected the LFWFs might not be utilized to find the full GPD. Consequently, if one requires that theoretical constraints should be consistently implemented or one simply aims for a more universal description of hadronic phenomena, such  popular ans\"atze are simply excluded.

Examples of effective two-body LFWFs overlap modeling were given in Ref.\  \cite{Muller:2014tqa}. Thereby, a ($t$-inde\-pendent) Regge behavior was implemented by means of a convolution with a spectator mass spectral  density. Such models are generically in agreement with experimental and phenomenological findings, however, various problems remain to be solved, such as mimicking a $t$-dependent Regge behavior that ensures positivity by construction and flexibility of the LFWF parametrization,    before on can develop a phenomenological LFWF framework that is suited for the description of experimental data.

\vspace{-0.3cm}

\begin{acknowledgements}
For discussions I am indebted to S.~Brodsky, D.~Chakrabarti, C.~Lorce, B. ~Pasquini, M.~Polyakov,  K.~Semenov-Tian-Shansky,  G.F.~de Teramond,   O.~Teryaev.
This work has been supported by the  Croatian Ministry of Science, Education and Sport (MSES) under the NEWFELPRO grant agreement
no.\ 54.
\end{acknowledgements}

\vspace{-0.4cm}


%
%

\end{document}